\documentstyle[epsfig]{aipproc}

\begin{document}

\font\fortssbx=cmssbx10 scaled \magstep2
\hbox to \hsize{{\fortssbx University of Wisconsin - Madison}
\hfill
\vtop{\hbox{\bf MADPH-01-1215}
      \hbox{January 2001}}}

\title{\vspace*{-1.5ex}
\mbox{${\bf 10}^{\bf 20}$eV Cosmic Ray and Particle Physics}
 with IceCube\thanks{Talk presented by F. Halzen at the {\it 1$^{st}$ International Workshop on Radio Detection of High Energy Particles} (RADHEP-2000), Los Angeles, California, November 2000}\vspace*{-1ex}}

\author{
J. Alvarez-Mu\~niz\thanks{Current Address: Bartol Research
Institute, University of Delaware, Newark, DE 19716} and F. Halzen}

\address{\vspace*{-2ex}
Physics Department, University of Wisconsin, Madison,
WI 53706, USA}

\maketitle

\vspace{-3ex}

\begin{abstract}\looseness=-1 
We show that a kilometer-scale neutrino observatory, though optimized  
for detecting neutrinos of TeV to PeV energy, can reveal the science  
associated with the enigmatic super-EeV radiation in the Universe.  
Speculations regarding its origin include heavy relics from the early  
Universe, particle interactions associated with the Greisen cutoff, and  
topological defects which are remnant cosmic structures associated  
with phase transitions in grand unified gauge theories. We show that it  
is a misconception that new instruments optimized to EeV energy can  
exclusively do this important science. Because kilometer-scale neutrino  
telescopes such as IceCube can reject the atmospheric neutrino background by  
identifying the very high energy of the signal events, they have  
sensitivity over the full solid angle, including the horizon where most  
of the signal is concentrated. This is critical because upgoing  
neutrino-induced muons, considered in previous calculations, are  
absorbed by the Earth. Previous calculations have underestimated the event rates of IceCube for EeV signals by over one order of magnitude.

\end{abstract}

\section{Introduction}

It is nothing less but exhilarating to contemplate future neutrino  
detectors reaching effective volumes of $10^{13}$\,tons and effective  
areas of $10^6$\,${\rm km^2}$ by exploiting totally novel detection methods  
such as radio, acoustic, atmospheric fluorescence and horizontal air  
shower techniques. This is at a time when we are operating a single  
neutrino telescope of only $10^3\mbox{--}10^5$\,${\rm m^2}$ effective area,  
depending on the science~\cite{amanda}. Its extension to a  
kilometer-scale neutrino observatory, IceCube, is still at the proposal  
stage. Neutrino detectors can be classified in four categories, which are delineated by the energy for which the instruments have been  
optimized:

\begin{enumerate}
\item {\bf MeV} detectors: for studying the sun and detecting supernovae,
\item {\bf GeV--TeV} DUMAND-class telescopes: possibly, the first  
instruments to look beyond the sun, but, more importantly, with  
sufficiently low threshold to demonstrate the novel techniques that use  
ice and water as a Cherenkov medium by detecting atmospheric  
neutrinos. AMANDA is the first in this category, others will be  
commissioned in Mediterranean waters,
\item {\bf TeV--PeV} kilometer-scale observatories such as  
IceCube\cite{icecube}. These represent, as far as we know, the best-buy  
for opening up the field of high energy neutrino astronomy,
\item {\bf EeV} detectors specializing in answering the mystifying  
questions raised by the existence of 100\,EeV cosmic rays, and the  
apparent absence of a Greisen cutoff.
\end{enumerate}

Although optimized in different energy bands, the missions of these  
instruments can overlap. In this talk we discuss the potential of  
IceCube to do the science envisaged for the projects of interest to  
this meeting, such as the radio observatories, the Auger air shower  
array and the space-based atmospheric fluorescence detector OWL. These  
discussions are important for two reasons: i) for exploring the full  
potential of a detector, possibly beyond the specific goals it was  
designed for, and ii) in order to avoid compromising the performance of  
an instrument by concentrating on science better done by others. While  
i) is obvious, ii) is important and often controversial. For instance,  
should one consider the study of oscillating atmospheric neutrinos,  
superbly performed with Superkamiokande-type detectors, when optimizing  
the performance a high energy neutrino telescope?

\section{IceCube}

As far as astronomy beyond the GeV signals of EGRET is concerned, the  
case for using neutrinos as messengers is compelling\cite{physrep}. Of all  
high-energy particles, only weakly interacting neutrinos can directly  
convey astronomical information from the edge of the universe and  
from deep inside the most cataclysmic high-energy processes. Copiously  
produced in high-energy collisions, travelling at the velocity of  
light, and undeflected by magnetic fields, neutrinos meet the basic  
requirements for astronomy. Their unique advantage arises from a  
fundamental property: they are affected only by the weakest of nature's  
forces (but for gravity) and are therefore essentially unabsorbed as  
they travel cosmological distances between their origin and us.

The first suggestions that kilometer-size neutrino telescopes were  
required to do the science originated with early estimates of the flux produced by the  
highest energy cosmic rays interacting with microwave photons. With  
time, and after consideration of the diverse scientific  
missions of astroparticle physics with high energy neutrinos, we have  
confirmed that the science does require construction of a  
kilometer-scale neutrino detector, and the challenge has therefore been  
one of technology. The only demonstrated solution is to use a ``natural''  
detector consisting of a thousand billion liters (a teraliter) of  
instrumented natural water or ice. After commissioning and operating  
the Antarctic Muon and Neutrino Detector Array (AMANDA), the AMANDA  
collaboration is ready to meet this challenge and has proposed to  
construct IceCube, a one-cubic-kilometer international high-energy  
neutrino observatory in the clear deep ice below the South Pole  
Station.

IceCube will be an array of 4,800 optical modules within a cubic  
kilometer of clear ice. Frozen into holes 2.4 kilometers deep, to be  
drilled by hot water, the uppermost optical modules will lie 1,400\,m  
below the surface. Simulations anchored to AMANDA data show that the direction of muon tracks can be reconstructed to 0.5  
degrees above 1\,PeV. IceCube will be capable of identifying neutrino type, or  
flavor, by mapping showers of Cherenkov light from electron and from  
tau neutrinos. Most important, it will measure neutrino energy. Energy  
resolution is critical, because there should be very little background  
from atmospheric neutrinos at energies above 100 TeV.

\section{E\lowercase{e}V Science}

In this talk we concentrate on super-EeV science such as topological  
defects, super-heavy relics and neutrinos associated with the Greisen  
cutoff. Their detection is usually not considered as a high priority in  
designing the architecture of neutrino telescopes.

It has been realized for some time that topological defects are  
unlikely to be the origin of the structure in the present Universe  
\cite{TDstructure}. Therefore the direct observation of their decay  
products, in the form of cosmic rays or high energy neutrinos, becomes  
the only way to search for these remnant structures from grand unified  
phase transitions\cite{TD}. This search represents an example of  
fundamental particle physics that can only be done with cosmic beams.  
We here point out that a kilometer-scale neutrino observatory, such as  
IceCube, has excellent discovery potential for topological defects. The  
instrument can identify the characteristic signatures in the energy  
and zenith angle distribution of the signal events. Our conclusions for  
topological defects extend to other physics associated with $10^{20}  
{\sim} 10^{24}$\,eV energies. 

To benchmark the performance of IceCube relative to OWL\cite{cline},  
chosen as an example of an instrument optimized to ${\sim} 10^2$\,EeV  
energy, we use the following theorized sources of super-EeV neutrinos:

\begin{itemize}

\item generic topological defects with grand-unified mass scale $M_X$  
of order $10^{15}$\,GeV and a particle decay spectrum consistent with  
all present observational constraints\cite{protheroe},

\item superheavy relics \cite{gelmini,sarkar}, which we normalize to the Z-burst  
scenario\cite{weiler} where the  
observed cosmic rays with ${\sim} 10^{20}$\,eV energy, and above, are  
locally produced by the interaction of super-energetic neutrinos with  
the cosmic neutrino background,

\item neutrinos produced by superheavy relics which themselves decay  
into the highest energy cosmic rays\cite{berez}, and

\item the flux of neutrinos produced in the interactions of cosmic  
rays with the microwave background\cite{steckerCMB}. This flux, which  
originally inspired the concept of a kilometer-scale neutrino detector,  
is mostly shown for comparison.

\end{itemize}

Our results are summarized in Table 1 where we compare the event  
rates for IceCube, discussed later, with those for OWL calculated in  
reference\cite{cline}. The conclusion is clear, while effective volume  
and area for OWL apparently exceed those of IceCube by many orders of  
magnitude, the events rates are comparable. This is a consequence of  
the duty cycle, reduced efficiency, and higher threshold of the OWL  
detector.

\begin{table}
\caption[]{Comparison of neutrino event rates for three
representative neutrino fluxes for OWL\cite{cline} and\break 
IceCube\cite{alvarez}.}
\tabcolsep.65em  
\def\arraystretch{1.3}
\centering\leavevmode
\begin{tabular}{lccc} 
& Volume & Eff. Area & Threshold\\
\cline{2-4}
 OWL & 10$^{13}$ ton & 10$^6$ km$^2$ & $3\times10^{19}$ eV\\
 IceCube & 10$^9$ ton & 1 km$^2$ & 10$^{15}$ eV$^*$\\
\hline
 & \multicolumn{3}{c}{Events per Year}\\[-1ex]
 & TD & $Z_{\rm burst}$ & $p\gamma_{2.7}$ \\
\cline{2-4}
OWL $\nu_e$& 16& 9& 5\\
IceCube $\nu_\mu$& 11& 30& 1.5\\
\hline
\end{tabular}
{\footnotesize$^*$actual threshold $\sim$100~GeV; requiring $>$1~PeV eliminates atmospheric
$\nu$ background.\hfill}
\end{table}

Cognoscenti will notice that the event rates claimed for IceCube are  
roughly two orders of magnitude larger than those found in the  
literature for a generic detector with 1\,km$^2$ effective area. The  
reason for this is simple. Unlike first-generation neutrino telescopes,  
IceCube can measure energy, and can therefore separate interesting  
high energy events from the background of lower energy atmospheric  
neutrinos by energy measurement. The instrument can identify  
high energy neutrinos over $4\pi$ solid angle, and not just in the lower  
hemisphere where they are identified by their penetration of  
the Earth, as is the case with AMANDA. This is of primary importance  
here because neutrinos produced, for instance by the decay of  
topological defects, have energies large enough to be efficiently  
absorbed by the Earth. The observed events are dominated by neutrinos  
interacting in the ice or atmosphere {\it above} the detector and near  
the horizon where the atmosphere alone represent a target density for  
converting neutrinos of $36\,{\rm kg/cm^2}$. This event rate typically  
dominates the one for up-going neutrinos by an order of magnitude. We  
will show that the zenith angle distribution of neutrinos associated  
with EeV signals form a striking signature for their extremely high  
energy origin.


\section{Neutrino events}

We calculate the neutrino event rates by convoluting the $\nu_\mu+\bar\nu_\mu$  
flux from the different sources considered in this talk, with the probability 
of detecting a muon produced in a  
muon-neutrino interaction in the Earth or atmosphere:

\begin{equation}
N_{\rm events}=2\pi~A_{\rm eff}~T~\int\int~{dN_\nu\over dE_\nu}(E_\nu)
P_{\nu\rightarrow\mu}(E_\nu,E_\mu {\rm (thresh)},\cos\theta_{\rm zenith})
~dE_\nu~d\cos\theta_{\rm zenith}
\end{equation}
where $T$ is the observation time and $\theta_{\rm zenith}$ the zenith 
angle. We assume an effective telescope area of  
$A_{\rm eff}=1~{\rm km^2}$,
a conservative assumption for the
very high energy neutrinos considered here.
It is important to notice that the probability ($P_{\nu\rightarrow\mu}$) 
of detecting a muon
with energy above a certain energy threshold $E_\mu$(threshold),
produced in a muon-neutrino interaction,
depends on the angle of incidence of the neutrinos. This is because the  
distance traveled by a muon cannot exceed the column density of matter available
for neutrino interaction, a condition not satisfied by very high energy  
neutrinos produced in the atmosphere. They are absorbed by the Earth and only  
produce neutrinos in the ice above, or in the atmosphere or Earth near the  
horizon. The event rates in which the muon arrives at the detector 
with an energy above $E_\mu\rm(threshold)=1~PeV$, where the atmospheric neutrino  background is negligible, are summarized in Table~2. We discuss them in more detail by introducing Figs.~1--5.

\begin{table}[h]
\caption[]{Neutrino event rates (per year per ${\rm km^2}$ in
$2\pi$ sr) in which the produced muon arrives
at the detector with an energy
above $E_\mu$(threshold)=1 PeV. Different neutrino sources have been
considered.
The topological defect models (TD) correspond to highest injection rates
$Q_0~({\rm ergs~cm^{-3}~s^{-1}})$
allowed in Fig.\,2 of \cite{protheroe}.
Also shown is the number of events from p-$\gamma_{\rm CMB}$ interactions
in which protons are propagated up to a maximum redshift $z_{\rm max}=2.2$ 
\cite{steckerCMB} and the number of neutrinos from the Waxman and Bahcall
limit on the diffuse flux from optically thin sources \cite{wblimit}.
The number of atmospheric background events above 1 PeV is also shown.
The second column corresponds to
downward going neutrinos (in $2\pi$ sr).
The third column gives the number of upward going
events (in $2\pi$ sr). We have taken
into account absorption in the Earth according to
reference \cite{gandhi}. IceCube will detect the sum of the event rates given  
in the last two columns.}
\centering\leavevmode
\def\arraystretch{1.25}
\begin{tabular}{ccc}
Model & $N_{\nu_\mu+\bar\nu_\mu}$ (downgoing) & $N_{\nu_\mu+\bar\nu_\mu}$
(upgoing)  \\ \hline\hline
TD, $M_X=10^{14}$ GeV, $Q_0=6.31\times 10^{-35}$, p=1~ & 11 & 1 \\
TD, $M_X=10^{14}$ GeV, $Q_0=6.31\times 10^{-35}$, p=2~ & 3 & 0.3 \\ 
TD, $M_X=10^{15}$ GeV, $Q_0=1.58\times 10^{-34}$, p=1~ & 9  & 1 \\
TD, $M_X=10^{15}$ GeV, $Q_0=1.12\times 10^{-34}$, p=2~ & 2 & 0.2 \\ 
Superheavy Relics Gelmini {\it et al.} \cite{gelmini} & 30 & $1.5\times 10^{-7}$ \\  
Superheavy Relics Berezinsky {\it et al.} \cite{berez} & 2 & 0.2 \\ 
Superheavy Relics Birkel {\it et al.} \cite{sarkar} & 1.5 & 0.3 \\ 
p-$\gamma_{\rm CMB}$ $(z_{\rm max}=2.2)$ \cite{steckerCMB} & 1.5 & $1.2\times 10^{-2}$ \\ 
W-B limit $2\times 10^{-8}~E^{-2}~{\rm (cm^2~s~sr~GeV)^{-1}}$ & 8.5 & 2 \\ \hline
Atmospheric background & $2.4\times 10^{-2}$ & $1.3\times 10^{-2}$ \\ \hline
\end{tabular}
\end{table}

\begin{figure}[t]
\begin{center}
\mbox{\epsfig{file=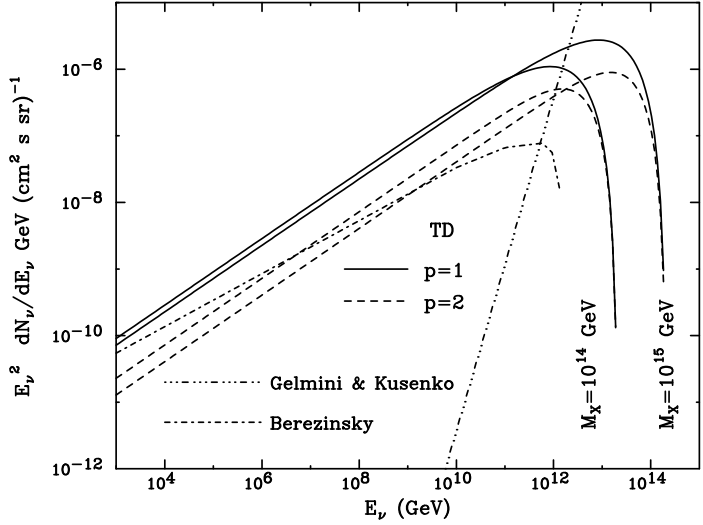,height=7cm}}
\end{center}
\caption{Maximal predictions of
$\nu_\mu+\bar\nu_\mu$ fluxes from topological defect models
by Protheroe and Stanev (p=1,2). Also
shown is the $\nu_\mu+\bar\nu_\mu$ from superheavy relic particles
by Gelmini and Kusenko and the flux by Berezinsky {\it et al.}.}
\end{figure}

Figure 1 shows the
$\nu_\mu+\bar\nu_\mu$ fluxes used in the calculations. We first calculate  
the event rates corresponding to the largest flux from topological  
defects \cite{protheroe} allowed by constraints imposed 
by the measured diffuse  
$\gamma$-ray
background in the vicinity of 100 MeV.  
The corresponding proton flux has been  
normalized to the observed cosmic ray spectrum
at $3\times 10^{20}$\,eV; see Fig.\,2 of reference \cite{protheroe}. 
Models with $p<1$ appear to
be ruled out \cite{sigl} and hence they are not considered in the
calculation. As an  
example of neutrino production by superheavy relic particles, we consider the  
model of  
Gelmini and Kusenko \cite{gelmini}. In Figs.\,2 and 3 we show the event rates  
as a function of neutrino energy. We assume a muon energy threshold
of 1 PeV. 
\begin{figure}
\begin{center}
\mbox{\epsfig{file=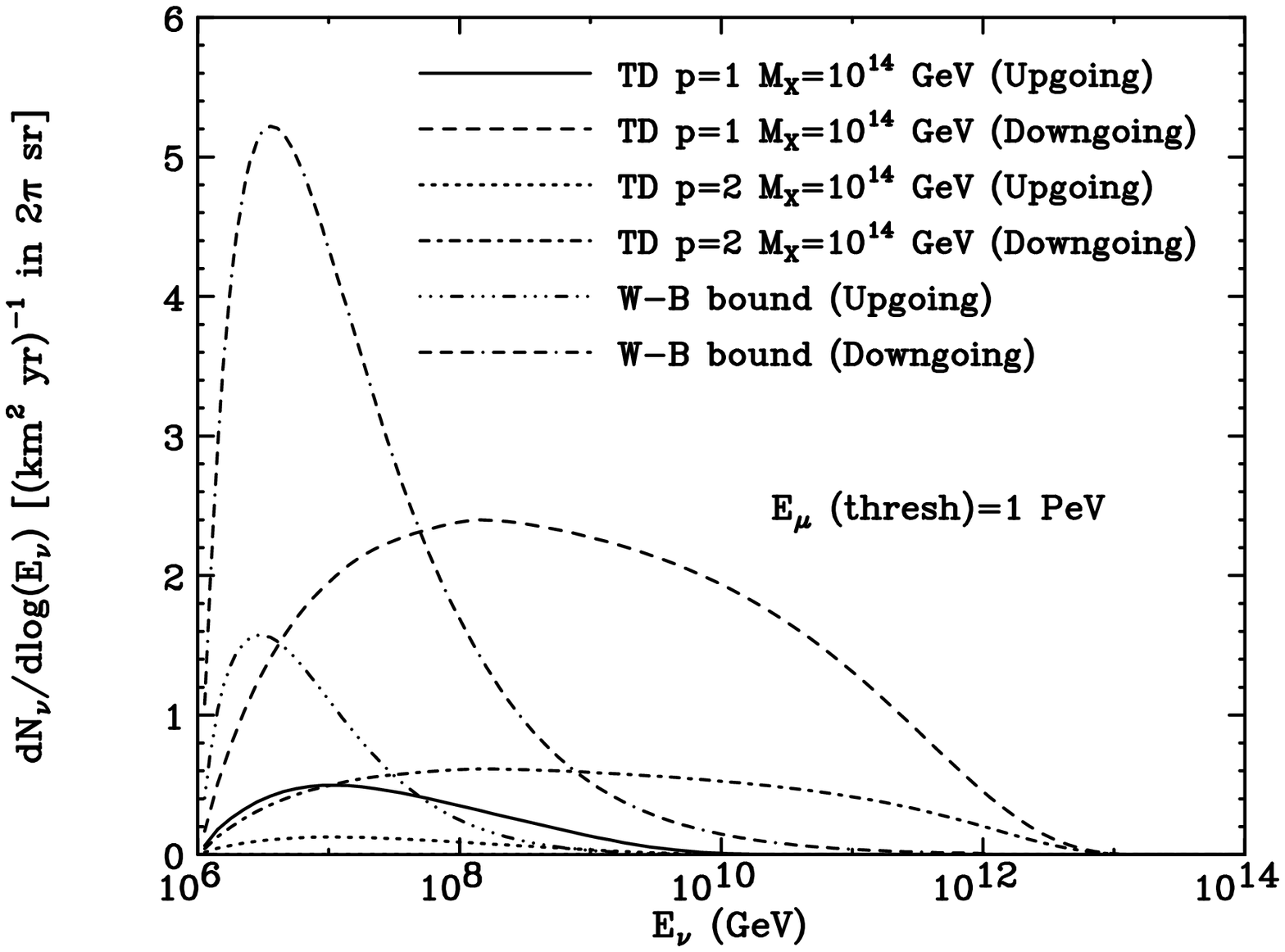,height=7cm}}
\end{center}
\caption{Differential $\nu_\mu+\bar\nu_\mu$ event rates in IceCube from
the topological defect fluxes in Fig.1.
The muon threshold is $E_\mu$(threshold)=1 PeV. 
We have separated the contribution from 
upgoing and downgoing events to stress the different behavior with 
energy. The event rate expected from the Waxman and Bahcall bound (see text)
is also shown for illustrative purposes. The rate due to atmospheric
neutrinos is negligible (see Table 2) and hence it is not plotted.}
\begin{center}
\mbox{\epsfig{file=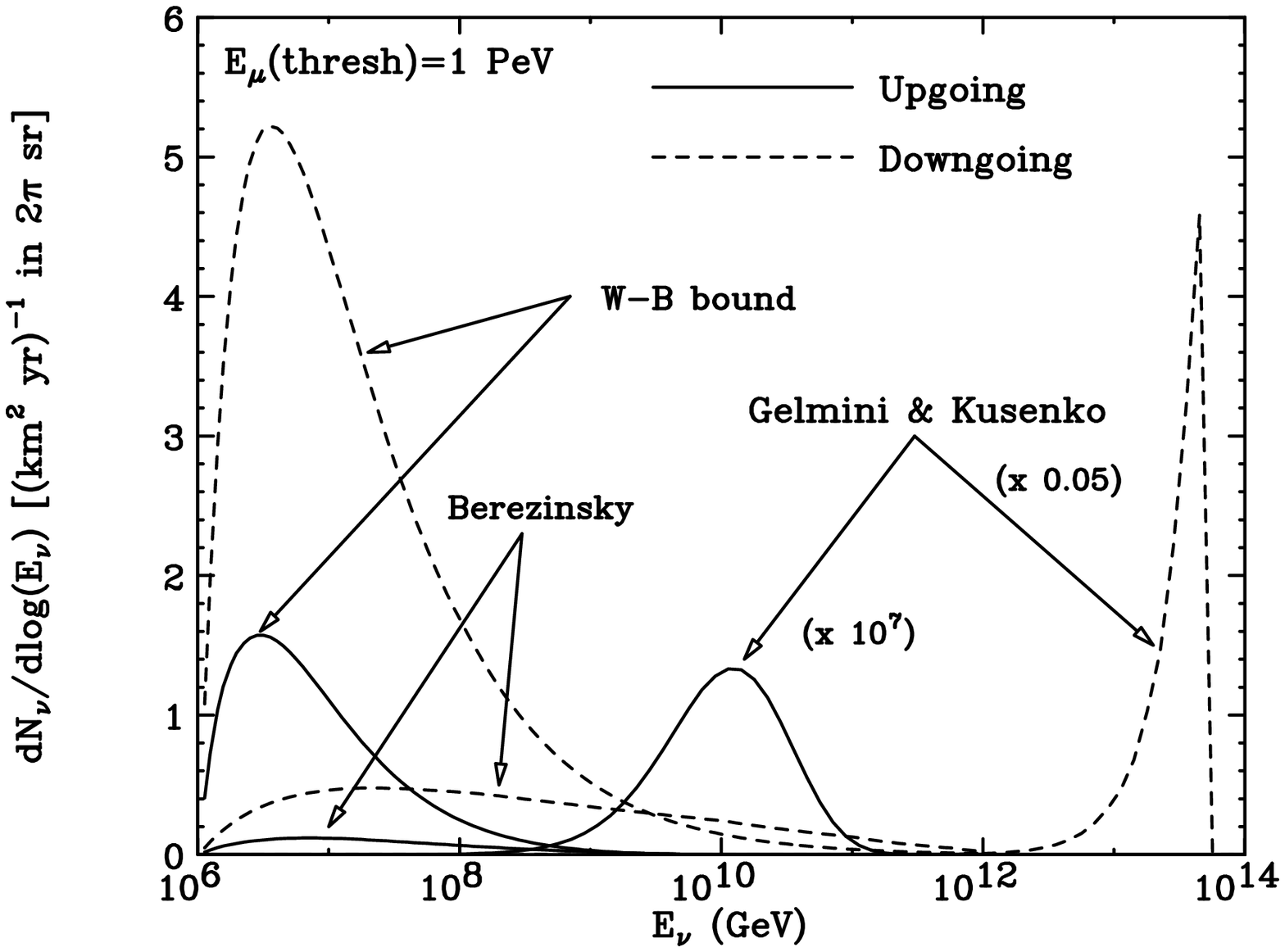,height=7cm}}
\end{center}
\caption{Differential $\nu_\mu+\bar\nu_\mu$ event rates in IceCube
from super-heavy relic particles. We have separated the contribution 
from upgoing and downgoing events to stress the different behavior with
energy.   
The muon threshold is $E_\mu$(threshold)=1 PeV. 
The event rate due to atmospheric 
neutrinos as well as the one expected from the Waxman and 
Bahcall bound (see text) is shown for illustrative purposes.
The rate due to atmospheric
neutrinos is negligible (see Table 2) and hence it is not plotted.}
\end{figure}
We also show in both plots the 
event rate corresponding to the Waxman and Bahcall ``bound" \cite{wblimit}. 
This bound represents the maximal
flux from astrophysical, optically thin sources, in which neutrinos are 
produced in $p$-$p$ or $p$-$\gamma$ collisions. The atmospheric
neutrino events are not shown since they are negligible above the 
muon energy threshold we are using. The area under the curves 
in both figures is equal to the number of events for each source.   
In Fig.\,4 we plot the
event rates in which the produced muon arrives at the detector
with an energy greater than $E_\mu$(threshold). In  
Fig.\,5 we finally present the angular distribution of the neutrino 
events for  
the different sources. The characteristic shape of  
the distribution reflects the opacity of the Earth to high energy neutrinos,  
typically  
above $\sim$100 TeV. The limited column density of matter in the atmosphere  
essentially reduces the rate of downgoing neutrinos to interactions in the  
1.5\,km of ice above the detector. The events are therefore concentrated  
near the horizontal direction corresponding to
zenith angles close to $90^\circ$. The neutrinos predicted by the model of  
Gelmini and Kusenko are so energetic that they are absorbed, even
in the horizontal direction as can be seen in Fig.\,5.

\begin{figure}[t]
\begin{center}
\mbox{\epsfig{file=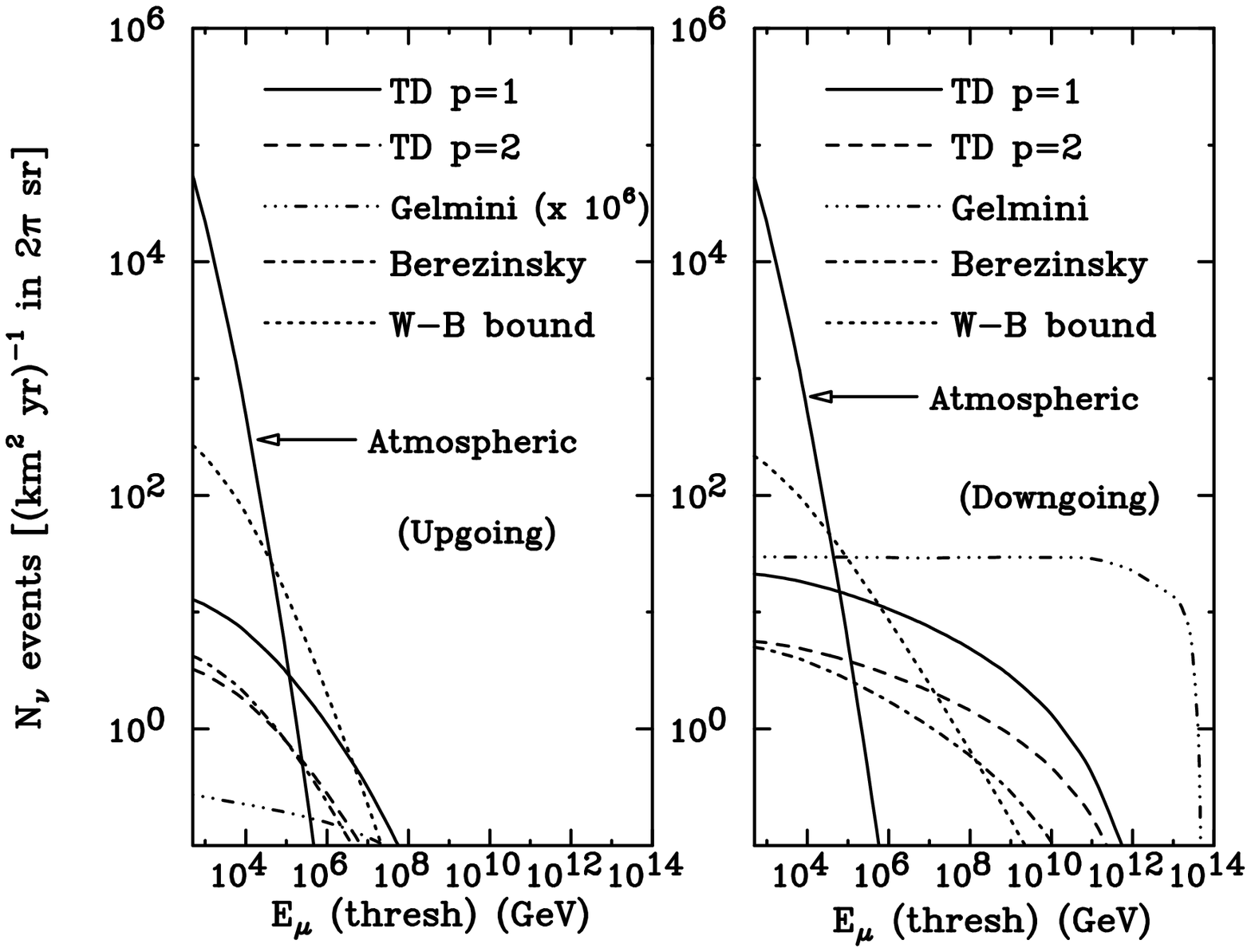,height=8.5cm}}
\end{center}
\caption{$\nu_\mu+\bar\nu_\mu$ event rates in IceCube from
the fluxes in Fig.1. The plot shows the number of events in which 
the produced muon arrives at the detector with an energy above $E_\mu$(thresh). 
Atmospheric neutrino events 
and the event rate expected from the Waxman and Bahcall upper bound
(see text) are also plotted. The topological defect (TD) models shown
(p=1 and p=2) correspond to $M_X=10^{14}$ GeV.
Upgoing and downgoing events are shown separately.} 
\end{figure}

\begin{figure}[t]
\begin{center}
\mbox{\epsfig{file=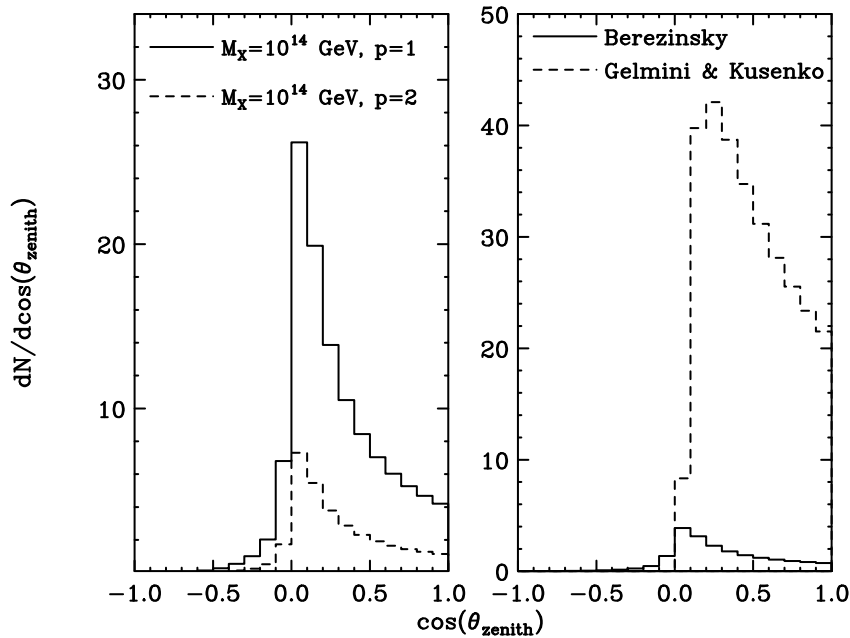,height=8.5cm}}
\end{center}
\caption{Zenith angle distribution of the  
$\nu_\mu+\bar\nu_\mu$ events in IceCube in which the produced muon arrives
at the detector with energy above 1 PeV.
Left: Topological defect models. Right: Superheavy relics. 
$\cos(\theta_{\rm zenith})=-1$
corresponds to vertical upgoing neutrinos, $\cos(\theta_{\rm zenith})=0$
to horizontal neutrinos and $\cos(\theta_{\rm zenith})=1$ to vertical
downgoing neutrinos. The detector
is located at a depth of 1.8 km in the ice.}
\end{figure}

Energy measurement is critical for achieving the sensitivity of the detector  
claimed. For muons, the energy resolution
of IceCube is anticipated to be $25\%$ in the logarithm of the energy,
possibly better. The detector is definitely able to determine energy to better than
an order of magnitude, sufficient for the separation of EeV signals from
atmospheric neutrinos with energies below 100 TeV. Notice that
one should also be able to identify electromagnetic showers initiated by
electron and tau-neutrinos. The energy response for showers is linear, and
expected to be better than 20\%. Such EeV events will be gold-plated,
unfortunately their fluxes are expected to be even lower. For instance
for the first TD model in Table 2 (p=1, $M_X=10^{14}$ GeV and 
$Q_0=6.31 \times 10^{-35}~{\rm ergs~cm^{-3}~s^{-1}}$), we expect 
$\sim 1$ contained shower 
per year per ${\rm km^2}$ above 1 PeV initiated in charged 
current interactions of $\nu_e+\bar\nu_e$. The corresponding number
for the Gelmini and Kusenko flux is $\sim 4~{\rm yr^{-1}~km^{-2}}$.

One should also worry about the fact that a very high energy muon may enter  
the detector with reduced energy because of energy losses. It can, in principle, become  
indistinguishable from a minimum ionizing muon of atmospheric origin \cite{gaisser}. We have 
accounted for the ionization as well as catastrophic muon energy losses 
which are incorporated in the calculation of the range of the muon. 
In the PeV regime region this energy reduction is roughly one 
order of magnitude, it should be less for  
the higher energies considered here. 

In conclusion, if the fluxes predicted by  
our sample of models for neutrino  
production in the super-EeV region are representative, they should be revealed  
by the IceCube observatory operated over several years.

\section*{Acknowledgements}
We thank J.J. Blanco-Pillado for making available to us his code to obtain
the neutrino fluxes from topological defects and E. Zas for helpful discussions.
This research was supported in part by the US Department of Energy under
grant DE-FG02-95ER40896 and in part by the University of Wisconsin
Research Committee with funds granted by the Wisconsin Alumni Research
Foundation.
The research activities of J.A. at Bartol Research Institute
are funded in part by NASA grant NAG5\_7009.

\end{document}